\documentstyle[prd,aps,twocolumn,tighten]{revtex}
\begin{document}
\title{Closed Lightlike Curves in Non-linear Electrodynamics}
\author{M. Novello$^1$\thanks{Electronic mail: 
\tt novello@lafex.cbpf.br}, V. A. De Lorenci$^2$\thanks{Electronic 
mail: \tt lorenci@cpd.efei.br}, E. Elbaz$^3$ and J. M. Salim$^1$}
\address{\mbox{}\\$^1$Centro Brasileiro de Pesquisas F\'{\i}sicas, \\
Rua Dr.\ Xavier Sigaud 150, Urca 22290-180 Rio de Janeiro, RJ -- Brazil} 
\address{$^2$Instituto de Ci\^encias -- 
Escola Federal de Engenharia de Itajub\'a, \\
Av.\ BPS 1303 Pinheirinho, 37500-000 Itajub\'a, MG -- Brazil}
\address{$^3$ Institut de Physique Nucl\'eaire de Lyon IN2P3-CNRS \\
Universit\'e Claude Bernard \\
43 Bd du 11 Novembre 1918, F-69622 Villeurbanne Cedex, France.}

\date{\today}
\renewcommand{\thefootnote}{\fnsymbol{footnote}}
\twocolumn[
\hsize\textwidth\columnwidth\hsize\csname@twocolumnfalse\endcsname 
\maketitle

\begin{abstract}
\hfill{\small\bf Abstract\hspace*{1.7em}}\hfill\smallskip
\par
\noindent
We show that non-linear electrodynamics may induce a photon to
follow a closed path in spacetime. We exhibit a specific case 
in which such closed lightlike curve (CLC) appears. 
\end{abstract}

\smallskip\mbox{}]
\footnotetext[1]{Electronic mail: \tt novello@lafex.cbpf.br}
\footnotetext[2]{Electronic mail: \tt lorenci@cpd.efei.br}
\renewcommand{\thefootnote}{\arabic{footnote}}

\newpage

\section{Introduction}
\subsection{Introductory Remarks}

One of the most elegant characteristics that singles out the
gravitational interaction consists in the possibility --turned into 
an actual formulation of gravity by Einstein --- of associating
gravitational phenomena to the metric 
structure of the spacetime. Such a geometrical view is 
impossible to be mimic by processes envolving other kinds of
forces. The main reason for this is that all 
other known 
forces do not show the universal property that is typical of gravity. 
Indeed, this was the main drawback that induced Einstein unified 
program to its failure. However 
an interesting analogy with the Einstein way of looking 
into some processes appeared recently, carrying a less
ambitious program but having a deep connection with it. 
It consists in a fresh method that ---taking into account 
such limitation of non-universality --- looks for special 
physical situations that allow an equivalent description in terms 
of an effective modification of the geometry of spacetime. In
other words: even if a given kind of force cannot be geometrized, 
one can discover some special (and non-trivial) situations in which 
a restricted geometrization is possible. It is clearly
understood that it is by no means a geometrization scheme in a broad 
sense but only a very limited one, although very 
useful, as we shall see. The 
interest on this arises, of course, from the possibility of applying 
such geometrization in a sufficiently large and meaningfull set of
examples. This is precisely the situation that one encounters in 
some interesting and diverse circumstances which has led 
to the claim that nongravitational processes can indeed 
simulate modifications of the geometry of spacetime.

Just to consider two cases 
that have been presented in a coherent and self-contained way we 
refer to \cite{novello99} and \cite{volovick} that deal respectivelly with:
\begin{itemize}
\item{The propagation of photons in non-linear electrodynamics;}
\item{Processes envolving certain properties of superfluid $^ {3}He.$}
\end{itemize}

In this paper we will deal only with the first case. 
Before going into this let
us make another rather general comment to clarify our work here. 
We will be concerned with the propagation of photons in a nonlinear 
electrodynamics in terms of a  modification of the metric of 
the spacetime. Such modification is nothing but an effective 
structure that yields an equivalent description of the photon paths. 
This should not be taken as an universal modification of the geometry
of the spacetime. We shall see however that such geometric tool is 
very powerful and allows the analysis of light propagation 
to be accomplished in a very simple way. Besides, with this method 
we can transpose part of the behavior of photons from the 
well-known combined Maxwell-Einstein framework 
to the nonlinear case of electrodynamics. An example of
this analogy will be presented in this paper. It concerns the 
possibility of the presence of photons closed paths in spacetime. We will 
see that the remarkable G\"odel analysis of the existence of 
closed timelike curves (CTC) in a rotating universe 
can be transposed to the case of photons in 
non-linear electrodynamics. The electromagnetic field generated by a charged 
string yields the possibility of the 
existence of closed lightlike curves (CLC) for the photons.

\subsection{Synopsis}

In the last years there has been an increasing interest 
on properties of the gravitational field that describe, 
within the context of General Relativity(GR), geometries that allow 
the appearance of closed timelike curves (CTC) (see, for instance 
\cite{frolov}, \cite{kim}, \cite{morris}, \cite{thorne1}, \cite{visser}). 
Such unusual geometries, exact solutions of the 
classical Einstein equations of GR, pose a  
thorough problem of compatibility in the realm of field theory 
--- just to quote one difficulty --- and it is a real 
challenge to deal with them. As an example, we can point out the case 
of traversable wormholes that 
allow the existence of two nonequivalent paths for the possible travel of 
a real observer to go from one point $P$ of spacetime to another 
point $Q$ inducing thus the existence of 
a closed path in spacetime. G\"odel cosmological solution is another case 
which also provides such kind of undesirable paths. The 
attraction for such geometries rests on the deep
understanding of the theory allowed by their analysis. In 
the present paper we show that similar paths can be generated in 
configurations of pure electromagnetic field in a non-linear regime.

In order to achieve such a proof we must first review some 
recent papers that show this hidden geometrical character of the 
photon propagation in nonlinear electrodynamics \cite{novello99}, 
\cite{dit}, \cite{elbaz}. 
We will recover such results in a very simple way in the next section.

\subsection{Definitions and notations}

We call the electromagnetic tensor $F_{\mu\nu}$, while its dual
$F_{\mu\nu}^*$ is 
\begin{equation}
F_{\alpha\beta}^* \doteq 
\frac{1}{2}\eta_{\alpha\beta}\mbox{}^{\mu\nu}F_{\mu\nu},
\label{0001}
\end{equation}
where $\eta_{\alpha\beta\mu\nu}$ is the completely antisymmetric 
Levi-Civita tensor; the Minkowski metric tensor is represented 
by its standard form $\eta^{\mu\nu}.$ 
The two invariants constructed with these tensors are defined as 
\begin{equation}
F \doteq F^{\mu\nu} \,F_{\mu\nu},
\end{equation}
\begin{equation}
G \doteq F^{\mu\nu} \,F_{\mu\nu}^*.
\end{equation}


Once the modifications of the linearity of electrodynamics 
which will be dealt here with 
do not break the gauge invariance of the theory, we can restrict our
analysis to the general form of the modified Lagrangian for 
electrodynamics, written as a functional of the above invariants. 
In the present paper we limit our analysis to the case in which 
the Lagrangian depends only on $F:$
\begin{equation}
L = L(F).
\label{s11}
\end{equation}

We denote by $L_F$ the derivative of the Lagrangian $L$ 
with respect to the invariant $F$; 
and similarly for the higher order derivatives.  
We are particularly interested in the derivation 
of the characteristic surfaces which guide the propagation 
of the field discontinuities.  

\section{The Method of the Effective Geometry}
We will make a short review of the Hadamard method in 
order to obtain the propagation equations for the discontinuities 
of the electromagnetic field. We limit our analysis to the case 
in which all modifications on the linear electrodynamics can be 
described by a Lagrangian $L$ as in Eq. (\ref{s11})

Let $\Sigma$ 
be a surface of discontinuity for the electromagnetic field.  
Following Hadamard \cite{novello99} we assume that 
the field itself is continuous when crossing $\Sigma$, 
while its first derivative presents a finite discontinuity.  
We accordingly set
\begin{equation}
\label{[F]}
[F_{\mu\nu}]_{\Sigma} = 0,
\label{gw1}
\end{equation} 
and   
\begin{equation}
\label{[Fnu]}
[\partial_{\lambda}\,F_{\mu\nu}]_{\Sigma} = f_{\mu\nu} k_{\lambda},
\label{gw2}
\end{equation} 
in which the symbol 
$$[J]_\Sigma \equiv \lim_{\delta\rightarrow 0^+} 
\left(J|_{\Sigma+\delta}-J|_{\Sigma-\delta}\right)$$ 
represents the discontinuity of the arbitrary function $J$ 
through the surface $\Sigma$ 
characterized by the equation $\Sigma(x^{\mu}) = constant$.  
The tensor $f_{\mu\nu}$ is called the discontinuity of the field, and 
\begin{equation}
k_\lambda=\partial_{\lambda}\,\Sigma
\label{000111}
\end{equation}
is the propagation vector.  

The equations of motion are 
\begin{equation}
 \label{field}
\partial_{\nu}\,\left(L_{F}F^{\mu\nu}\right) = 0.
\end{equation}
Following the definitions and procedure  presented 
above one gets from the discontinuity of the equation of motion 
Eq. (\ref{field}):

\begin{equation}
L_{F}f^{\mu\nu}\, k_{\nu} + 2 L_{FF} \,\xi F^{\mu\nu} k_{\nu} = 0,
\label{gw3}
\end{equation} 
where $\xi$ is defined by 
\begin{equation}
\xi  \equiv  F^{\alpha\beta} \, f_{\alpha\beta}.
\end{equation}
The cyclic identity yields
\begin{equation}
f_{\mu\nu} k_{\lambda} + f_{\nu\lambda} k_{\mu} + 
f_{\lambda\mu} k_{\nu} = 0.
\label{gw4}
\end{equation}
Multiplying this equation by $k^{\lambda}\, F^{\mu\nu}$ gives 
\begin{equation}
\xi k_{\nu} \,k_{\mu} \gamma^{\mu\nu} + 2 \,F^{\mu\nu} 
f_{\nu\lambda} k^{\lambda} \, k_{\mu} = 0, 
\end{equation}
in which $\gamma_{\mu\nu}$ is the Minkowski metric tensor written in 
an arbitrary coordinate system.
From the Eq. (\ref{gw3}) it results:
\begin{equation}
f_{\mu\nu} \, k^{\nu} = - \,2\, \frac{L_{FF}}{L_{F}} \, \xi 
F_{\mu\nu} \, k^{\nu}.
\end{equation}

After some algebraic manipulations the equation of propagation of the
disturbances is obtained:
\begin{equation}
\left\{\gamma^{\mu\nu} + \Lambda^{\mu\nu} \right\} k_{\mu} k_{\nu} = 0
\label{gww4}
\end{equation}
in which the quantity $\Lambda^{\mu\nu}$ is 
\begin{equation}
\Lambda^{\mu\nu} \equiv - 4\, \frac{L_{FF}}{L_{F}} \,
F^{\mu\alpha} \,F_{\alpha}\mbox{}^{\nu}.
\label{lam1}
\end{equation}

It then follows that the photon path is kinematically described 
by 
\begin{equation}
\label{g2ef}
g^{\mu\nu}\,k_{\mu}\,k_{\nu} = 0,
\end{equation}
where the effective metric $g^{\mu\nu}$ is given by 
\begin{equation}
g^{\mu\nu} = L_F\gamma^{\mu\nu} - 4\,L_{FF}\, F^{\mu}\mbox{}_{\lambda}
F^{\lambda\nu.} 
\label{geral}
\end{equation}

Furthermore, once the wave vector $k_{\alpha}$ is a gradient, the 
photon path is a geodesic in the effective geometry 
\cite{novello99}. We re-obtained then the remarkable result that 
the discontinuities of the electromagnetic field 
in a nonlinear electrodynamics propagates along 
{\bf null geodesics} of an effective geometry 
which depends on the properties of the background
field\footnote{The proof that the path is indeed a geodesics is given
in the appendix B.}.

From the general expression of the energy-momentum tensor 
for an electromagnetic theory $L=L(F)$ we have
\begin{equation}
T_{\mu\nu} = - 4 L_{F}\, {F_{\mu}}^{\alpha} \, F_{\alpha\nu} - 
L \,\gamma_{\mu\nu}. 
\protect\label{111}
\end{equation}

We can then re-write the effective geometry in a more 
appealing form in terms of the energy momentum tensor. We 
obtain, using Eq. (\ref{111}) into Eq. (\ref{geral})
\begin{equation}
g^{\mu\nu} = {\cal M}\,\gamma^{\mu\nu} + {\cal N} \,T^{\mu\nu},
\label{AB}
\end{equation}
where the functions ${\cal M}$ and ${\cal N}$ are given by
\begin{equation}
{\cal M}  =  L_F +  \frac{L\,L_{FF}}{L_{F}},
\label{A}
\end{equation}
\begin{equation}
{\cal N}  =  \frac{L_{FF}}{L_{F}}
\label{B}
\end{equation}
As a consequence of this, the 
Minkowskian norm of the propagation vector $k_\mu$ reads 
\begin{equation}
\gamma^{\mu\nu} k_{\mu}\,k_{\nu} = - \frac{{\cal N}}{{\cal M}}
T^{\mu\nu}k_{\mu}k_{\nu}.
\end{equation}

\section{Charged String}

The physical system we will analyse consists in a (idealized infinitely 
long) thin charged cylinder\footnote{We neglect all
gravitational effects once there is no substantial difference 
introduced by gravity, as far as the phenomenon we are interested to 
exhibit here is concerned.}. The flat Minkowskian background 
geometry written in a 
$(t, r, \varphi, z)$ coordinate system takes the
form

\begin{equation}
ds^2 = dt^2 - dr^2 + 2 h_{0} d\varphi\, dt + g(r)\,d\varphi^{2} - 
dz^2
\protect\label{S1}
\end{equation}
where $g(r) = h_{0}^{2} - \omega^{2}\, r^2.$ 
Although locally such geometry is Minkowskian, it can
be associated to a spinning string, due to its global properties. We
shall see that this has no effect on our analysis, once 
CLC\rq s do not exist in the Maxwell-Einstein 
theory but only in the case of nonlinear electrodynamics. This 
shows that our example of CLC is not a gravitational effect.
Furthermore, one could set $\omega = 1$ in all calculations without 
any qualitative changing in our result. In order to avoid any 
artificial trouble with causality we will limit 
the range of the coordinate $r$ to be strictly larger than $r_{0},$ that 
is, $r > r_{0}$ where 
$r_{0}^{2} = h_{0}^{2}/\omega^{2}.$ In this domain of validity, this 
system is regular and well defined \cite{marta}.

From the symmetry properties of this system, the 
unique non-null component of the electric field is 
$F_{01} = E(r)$. In this case\footnote{In order to 
avoid any difficulty with the coordinates we take the 
radius of the charged string to be greater that the value $r_{0}.$}, 
the equation of motion reduces to
\begin{equation}
L_{F}\, E = \frac{Q}{r}
\protect\label{S2}
\end{equation}
where $Q$ is a constant. We are interested here in the 
analysis of the propagation of electromagnetic waves in such 
background. Following our previous treatment the photons 
propagate as if the metric structure of spacetime were changed 
into an effective Riemannian geometry. 
From Eq.(\ref{geral}) we obtain the components of the effective metric. The 
non-vanishing covariant components\footnote{See the appendix A.} are:
\begin{equation}
g_{tt} = \frac{\omega^{2}\,r^{2}}{h_{0}^{2} + \Psi\,\omega^{2}\, r^{2}}
\protect\label{S3}
\end{equation}
\begin{equation}
g_{t\varphi} = h_{0} \, g_{tt}
\protect\label{S4}
\end{equation}\begin{equation}
g_{rr} = \frac{1}{\Lambda}
\protect\label{S5}
\end{equation}
\begin{equation}
g_{\varphi\varphi} = \,-\,\Psi\,\omega^{2}\,r^{2} \,g_{tt}
\protect\label{S6}
\end{equation}
\begin{equation}
g_{zz} = -\,1,
\protect\label{S7}
\end{equation}
where $\Psi$ and $\Lambda$ are 
\begin{equation}
\Psi = 1 - \left(\frac{h_{0}}{\omega\,r}\right)^{2} 
- 4\,\frac{L_{FF}\,E^{2}}{L_{F}}
\protect\label{zz1}
\end{equation}
\begin{equation}
\Lambda = -\,1 - 4\,\frac{L_{FF}\,E^{2}}{L_{F}}.
\protect\label{zz2}
\end{equation}

The photon paths are null geodesics in such modified geometry.  
Let us consider the curve defined by the equations 
$ t = constant,$ $r = constant,$ and $z = constant.$ Along such a curve
the element of length reduces to 
 
\begin{equation}
ds^{2}_{eff} = -\,\Psi\,\omega^{4}\,r^{4} \,
\left(\frac{1}{h_{0}^{2} + \Psi\,\omega^{2}\, r^{2}}\right)\,d\varphi^{2}.
\protect\label{zz3}
\end{equation}

Thus, for the photon to follow such a path the radius $r = r_{c}$
must be such that $\Psi(r_{c}) = 0.$  

Let us
emphasize that the possibility of the presence of CLC\rq s depends
crucially on the non-linearity of the electromagnetic field. Indeed,
it is a direct consequence of the form of the above geometry that 
$\Psi$ can vanish only if $L_{FF}$ is different from 
zero. In the linear Maxwell electrodynamics this phenomenon is
forbidden\footnote{Let us remark that a similar analysis on the photon path 
can be made for Maxwell theory in a non-linear dielectric medium. 
This is a direct consequence of the propagation equations in a 
non-linear dielectric medium as it was shown in \cite{novello99}.} . 
Thus we are allowed to claim that it is 
a new property which depends on the non-linearity of the 
electromagnetic process. 
This is a general formalism valid for arbitrary form of the 
Lagrangian. We now turn to a specific example in which such situation occurs. 

\subsection{A toy model}
We set for the nonlinear Lagrangian the form\footnote{This form 
is very similar to Born-Infeld (BI) model. However, the sign 
in the field term inside the 
square-root is opposite to that used in BI Lagrangian. 
This makes a crucial difference in what 
concerns the appearance of CLC, as shown in the text.}:

\begin{equation}
L = \frac{b^{2}}{2} \,\left(\,\sqrt{1 - \frac{F}{b^{2}}} - 1\,\right)
\protect\label{S8}
\end{equation}
in which $b$ is an arbitrary constant.

A solution of the equation for the electric field yields
\begin{equation}
E = 4\,b\,Q\,\left(b^{2}\,r^{2} - 32\, 
Q^{2}\right)^{-\,\frac{1}{2}}.
\label{s9}
\end{equation}
The field must be defined for any value of $r$ larger than $r_{0}.$
Thus, the minimum value of the radius $r_{min}$ that follows 
from this expression must be larger than 
$\left(\frac{h_{o}}{\omega} \right)^{2},$ yielding a compromise
between the constants. Using this value on the expression of the 
effective metric gives
\begin{eqnarray}
\label{s10}
ds^{2}_{eff} = \frac{r^2}{r^2 - l^2}\, dt^2 + 
\frac{2\,h_{0}\,r^2}{r^2 - l^2}\, dt\,d\varphi &-& \nonumber \\ 
\frac{r^2}{r^2 + l^2} \,dr^2 - \omega^{2}\,r^2\,\frac{r^2 - l^2 - 
\frac{h_{0}^2}{\omega^2}}{r^2 - l^2} \,d\varphi^2 - dz^2,
\end{eqnarray}
where $l^{2} \equiv \frac{32\,Q^{2}}{b^{2}}.$ 
The value for which $\Psi$ vanishes is given by 
$$r_{c}^{2} = \left(\frac{h_{0}}{\omega}\right)^{2} + 
\frac{32\,Q^{2}}{b^{2}}.$$ 
For $r = r_{c}$ the photon follows a closed spacetime path.

\section{Final Comments}
It has been known from more than half a century that gravitational 
processes allows the existence of closed paths in spacetime. This led
to the belief that this strange situation occurs uniquely under the 
effect of gravity. In the present paper we have shown that this is
not the case. Indeed, we show here that photons can follow 
closed paths (CLC) 
due to electromagnetic forces in a non-linear regime. We presented 
an specific example of a theory in which CLC exists. In the limit
case where the non-linearities are neglected the presence of CLC 
is no more possible. Thus we are allowed to claim that this 
new property depends crucially on the non-linearity of the 
electromagnetic process and it is not possible to exist in Maxwell theory. 

This shows that the existence of CLC is {\bf not} an exclusive 
property of gravitational interaction: 
it can exists also in pure electromagnetic processes 
depending on the non-linearities of the background field. 
The existence of such CLC in both gravitational and electromagnetic 
processes asks for a deep review of the causal structure 
displayed by the photon path.

\section{Appendix A: The inverse metric}

In the case in which the Lagrangian depends only on the 
invariant $F$ the effective geometry takes the form
\begin{equation}
g^{\mu\nu} = L_F\gamma^{\mu\nu} - 4\, L_{FF}\, F^{\mu}\mbox{}_{\lambda}
F^{\lambda\nu} 
\label{apa1}
\end{equation}
The inverse metric, the covariant tensor $g_{\mu\nu}$  defined by 
$$g^{\mu\nu}\,g_{\nu\alpha} = \delta^{\mu}_{\alpha}$$
can be easily evaluated by taking into account the identities
$$F^{\mu}\mbox{}_{\lambda}\, F^{\lambda\nu} - 
F^{*\,\mu}\mbox{}_{\lambda}\, F^{*\,\lambda\nu} = -\,\frac{1}{2}\,F 
\gamma^{\mu\nu},$$
and
$$F^{*\,\mu}\mbox{}_{\lambda}\, F^{\lambda\nu} = -\,\frac{G}{4}\, 
\gamma^{\mu\nu}.$$
A direct manipulation yields the result
\begin{equation}
g_{\mu\nu} = A\,\gamma_{\mu\nu} + B\,F_{\mu}\mbox{}^{\lambda}\, F_{\lambda\nu}
\label{apa2}
\end{equation}
where $A$ and $B$ are 
$$A = \frac{1}{R}\,\left(L_{F} + 2\,F\,L_{FF}\right),$$ 
$$B = \frac{4\,L_{FF}}{R},$$
and $R$ is defined by 
$$R = L_{F}{}^{2} + L_{FF}\,\left(2\,F\,L_{F} - G^{2}\right).$$

\section{Appendix B: The Effective Null Geodesics}

The geometrical relevance of the effective geometry (\ref{geral}) 
goes beyond its immediate definition.  Indeed, we will 
demonstrate here that the integral curves of the vector $k_\nu$ 
({\em i.e.}, the trajectories of such nonlinear {\em photons}) 
are in fact geodesics.  
In order to achieve this result it will be required 
an underlying Riemannian structure for the manifold 
associated with the effective geometry.  In other words, 
this means a set of Levi-Civita connection coefficients 
$\Gamma^\alpha\mbox{}_{\mu\nu}=\Gamma^\alpha\mbox{}_{\nu\mu}$, 
by means of which there exists a covariant differential operator 
$\nabla_\lambda$ (the {\em covariant derivative}) such that 
\begin{equation}
\label{Riemann}
\nabla_\lambda g^{\mu\nu}\equiv g^{\mu\nu}\mbox{}_{;\,\lambda} 
\equiv g^{\mu\nu}\mbox{}_{,\,\lambda} + 
\Gamma^\mu\mbox{}_{\sigma\lambda}g^{\sigma\nu} + 
\Gamma^\nu\mbox{}_{\sigma\lambda}g^{\sigma\mu}=0.  
\end{equation}
From (\ref{Riemann}) it follows that 
the effective connection coefficients are completely determined 
from the effective geometry by the usual Christoffel formula.  

Contracting (\ref{Riemann}) with $k_\mu k_\nu$ results 
\begin{equation}
\label{N15}
k_\mu k_\nu g^{\mu\nu}\mbox{}_{,\,\lambda} = 
-2k_\mu k_\nu\Gamma^\mu\mbox{}_{\sigma\lambda}g^{\sigma\nu}.
\end{equation}
Differentiating (\ref{g2ef}) and remembering $g^{\mu\nu}=g^{\nu\mu}$ 
one gets
\begin{equation}
\label{N16}
2k_{\mu,\,\lambda}k_\nu g^{\mu\nu} + 
k_\mu k_\nu g^{\mu\nu}\mbox{}_{,\,\lambda} = 0.
\end{equation}
Inserting (\ref{N15}) 
for the last term on the left hand side of (\ref{N16}) we obtain 
\begin{equation}
\label{N17}
g^{\mu\nu}k_{\mu,\,\lambda}k_\nu - 
g^{\sigma\nu}\Gamma^\mu\mbox{}_{\sigma\lambda}k_\mu k_\nu = 0.
\end{equation}
Relabeling contracted indices we can rewrite (\ref{N17}) as 
\begin{equation}
\label{N18}
g^{\mu\nu}k_{\mu;\,\lambda}k_\nu \equiv 
g^{\mu\nu}\left[k_{\mu,\,\lambda} - 
\Gamma^\sigma\mbox{}_{\mu\lambda}k_\sigma\right]k_\nu = 0.
\end{equation}
Now, as the propagation vector $k_\mu=\Sigma_{,\,\mu}$ 
is an exact gradient one can write 
$k_{\mu;\,\lambda}=k_{\lambda;\,\mu}$.  
With this identity and defining $k^\nu\equiv g^{\mu\nu}k_\mu$ 
equation (\ref{N18}) reads 
\begin{equation}
\label{geodesic}
k_{\mu;\,\lambda}k^\lambda = 0,
\end{equation}
which states that $k_\mu$ is a geodesic vector.  
Remembering it is also a null vector 
(with respect to the effective geometry $g^{\mu\nu}$) 
its integral curves are thus null geodesics.  
We can restate our previous demonstration as:
\begin{quote}

The discontinuities of the electromagnetic field 
in a nonlinear electrodynamics propagates along 
{\bf null geodesics} of an effective geometry 
which depends on the properties of the background field.
\end{quote}

Thus, photons follow geodesics in an effective geometry. Since the
photon does not have electric charge, the force that acts on it 
in a non-linear electromagnetic field has a distinct character than
that of the Lorentz force. Indeed, from the geodesic 
equation, the electromagnetic field 
acts on the photon by means of a force given by 
\begin{equation}
f^{\alpha} = - \Delta^{\alpha}{}_{\mu\nu}\, k^{\mu}\,k^{\nu}
\label{apb1}
\end{equation}
in which the quantity $\Delta^{\alpha}{}_{\mu\nu}$ can be displayed 
in terms of the effective geometry $g^{\mu\nu}$ and its inverse 
$g_{\mu\nu}$ given above by the Christoffel form.  
$$\Delta^{\alpha}{}_{\mu\nu} = \frac{1}{2}\, ( L_{F}\,
\gamma^{\alpha\beta} + \Phi^{\alpha\beta} ) 
\left(\partial_{\nu}\,g_{\beta\mu} +  \partial_{\mu}
\,g_{\beta\nu}  - \partial_{\beta}\,g_{\mu\nu}\right)$$
where $\Phi^{\alpha\beta} \equiv 
-4\, L_{FF}\, F^{\alpha}\mbox{}_{\lambda} F^{\lambda\beta}.$ 
Hence, it follows that the net effect of the force that the field 
exerts on the photon has very similar properties as 
the gravitational force. This is precisely the reason that allows 
us to interpret the action of the electromagnetic field 
on the photon to be a mimic of the behavior of a massless particle 
in a gravitational field.

\end{document}